\documentclass[useAMS,usenatbib]{mn2e}
\usepackage{graphicx,color,epsfig}

\newcommand{\be}{\begin{equation}}
\newcommand{\ee}{\end{equation}}

\newcommand{\apj}{ApJ}

\newcommand{\mnras}{MNRAS}
\newcommand{\aap}{A\&A}

\newcommand{\apjl}{ApJL}

\newcommand{\nat}{Nature}

\newcommand{\icarus}{ICARUS}

\def\ltsima{$\; \buildrel < \over \sim \;$}
\def\simlt{\lower.5ex\hbox{\ltsima}}
\def\gtsima{$\; \buildrel > \over \sim \;$}
\def\simgt{\lower.5ex\hbox{\gtsima}}

\def\msun{{\,{\rm M}_\odot}}

\newcommand\mearth{{\,{\rm M}_{\oplus}}}

\def\del#1{{}}

\title[Tidal Downsizing and comet composition]{The Tidal Downsizing hypothesis for planet
  formation and the composition of Solar System comets.}

\author[S. Nayakshin, S.-H. Cha and J. Bridges]{Sergei Nayakshin$^{1,*}$, Seung-Hoon Cha$^1$ and John Bridges$^2$\\ 
$^1$ Department of Physics \& Astronomy,
  University of Leicester, Leicester, LE1 7RH, UK\\
$^2$ Space Research Centre, Department of Physics \& Astronomy,
  University of Leicester, Leicester, LE1 7RH, UK\\
$^*${E-mail:~} {\rm Sergei.Nayakshin@astro.le.ac.uk}}

\begin{document}

\date{Received}

\pagerange{\pageref{firstpage}--\pageref{lastpage}} \pubyear{2008}

\maketitle

\label{firstpage}

\begin{abstract}
Comets are believed to be born in the outer Solar System where the temperature
is assumed to have never exceeded $T \sim 100$ K. Surprisingly, observations
and samples of cometary dust particles returned to Earth showed that they are
in fact made of a mix of ices, as expected, but also of materials forged at
high-temperatures ($T\sim 1500$ K).  We propose a radically new view regarding
the origin of the high-temperature processed materials in comets, based on the
recent ``Tidal Downsizing'' (TD) hypothesis for planet formation. In the
latter, the outer proto-planetary disc is gravitationally unstable and forms
massive giant planet embryos (GEs). These hot ($T\sim $ hundreds to $2,000$ K)
and dense regions, immersed in the background cold and low density disc, are
eventually disrupted. We propose that both planets and the high-T materials in
comets are synthesised inside the GEs. Disruption of GEs separates planets and
small solids as the latter are ``frozen-in'' into gas and are peeled off
together with it. These small solids are then mixed with the ambient cold disc
containing ices before being incorporated into comets. Several predictions of
this picture may be testable with future observations of the Solar System and
exoplanets.
\end{abstract}


\section{Introduction}\label{intro}

Comets are icy bodies $\simgt$ km across that leave spectacular tails of
material (dust) when ices are vaporised. Comets contain some of the most
pristine materials from the dawn of the Solar System, and offer vital clues
about its formation process.  The composition of comets is confusingly
diverse.  Some of the materials found in cometary nuclei, e.g., amorphous
water and ammonia ices, have never \citep{KawakitaEtal04} experienced
temperatures above $\sim 30 - 150$ K, confirming their formation very far out,
probably around the present day orbits of Uranus and Neptune. However, the
mass fraction of crystalline silicates in the comae of the short period comet
81P/Wild 2, and in the ejecta of comet 9P/Tempel 1 is tantalisingly high
\citep{ZolenskyEtal06}, perhaps as high as \citep{WestphalEtal09} $\psi \sim
0.5 -0.65$. This is surprising as some crystalline silicates such as olivine
require temperatures in excess of 1000 K to make \citep{WoodenEtal07},
although not all crystalline silicates form at high temperature.

In the ``Core Accretion'' (CA) paradigm of planet formation
\citep{Safronov72,PollackEtal96,IdaLin08}, the outer disc is a rather
uninteresting and cold place where planet formation is not very likely as the
solid core formation time scales are long \citep{Safronov69,Rafikov10}.  The
temperature outside $R\sim 10$ AU is generally expected to remain below 100
K. Therefore, in the context of CA model, the presence of materials made at $T
\simgt 1000$ K in comets strongly suggests \citep{WoodenEtal05} a radial
transport of hight-T grains from the inner $R\simlt 1$ AU regions into the
outer, $R\simgt 10-30$ AU regions. Detailed models of the process
\citep{Gail01,HA10} show numerous constraints necessary to satisfy in order to
yield a significant enough outward transfer of solids.

Here we show that a set of recent ideas
\citep{BoleyEtal10,Nayakshin10a,Nayakshin10b,Nayakshin10c}, proposed as an
alternative to the CA model for planet formation, as a by-product may
naturally explain the otherwise puzzling composition of comets. The defining
difference of the model from the CA scenario, as far as comet formation is
concerned, is the non-unique radius-temperature relation in the disc. 

In the TD model, in a stark contrast to the CA picture, the outer disc is the
most important region for planet formation, as it is the birthplace of the
giant planet embryos. These massive ($\sim 10$ Jupiter masses) planet-forming
gas clumps are very hot and dense not due to being close to the parent star or
viscous disc heating, but simply due to contraction of the clumps. The clumps
are in fact undermassive isolated ``first cores'' -- embryos of stars
\citep{Larson69} -- that are not destined to develop into a low mass star due
to the imposing presence of the parent star \citep{Nayakshin10c} that anchors
the protostellar disc. It should thus not be surprising that these clumps
manage to become as hot as $\sim 1000$ K all on their own, at arbitrary
distances from the parent star.

The first cores (gas clumps) are excellent sites for grain growth
\citep{Nayakshin10a,Nayakshin10b} and thermal processing of solid
materials. Inside of these hot gaseous ``ovens'', chemical compounds can be
baked into materials not normally expected to form at tens to hundreds of
AU. Furthermore, a vital part of the TD model is the eventual disruption of
gas embryos which release the planets back into the ``ambient'' disc. This
disruption process, as we argue below, also releases the smaller thermally
reprocessed solids back into the disc. The thermally reprocessed materials can
then be rapidly mixed with the cold materials. As this is an {\em in situ}
model, no outward transport of solids is required.

\section{The Tidal Downsizing hypothesis}\label{sec:TD}

The TD hypothesis is a new combination of earlier well known ideas and
contains four important stages \citep[as illustrated in Figure 1
  of][]{Nayakshin10e}:

\begin{itemize}

\item[(1)] Formation of gas clumps (which we also call giant planet embryos;
GEs). As the protoplanetary disc cannot fragment inside $R \sim 50$ AU %
\citep{Rafikov05,BoleyEtal06}, GEs are formed at somewhat larger radii. The
mass of the clumps is estimated at $M_{\mathrm{GE}} \sim 10 M_J$ (10 Jupiter
masses) \citep{BoleyEtal10,Nayakshin10a}; they are intially fluffy and cool (%
$T\sim 100$ K), but contract with time and become much hotter %
\citep{Nayakshin10a}. \vskip 0.2 cm

\item[(2)] Inward radial migration of the clumps due to gravitational
  interactions with the surrounding gas
  disc \citep{GoldreichTremaine80,Lin96,VB10,BoleyEtal10,ChaNayakshin10}.
\vskip 0.2 cm

\item[(3)] Grain growth and sedimentation inside the clumps
  \citep{McCrea60,McCreaWilliams65,Boss98,BossEtal02}. If the clump
  temperature remains below $1400-2000 $K, massive terrestrial planet cores
  may form \citep{Nayakshin10b}, with masses up to the total high Z element
  content of the clump (e.g., $\sim 60 \mearth$ for a Solar metalicity clump
  of total mass $10 M_J$).
\vskip 0.2 cm

\item[(4)] A disruption of GEs in the inner few AU due to tidal
  forces \citep{McCreaWilliams65,BoleyEtal10,Nayakshin10c} or due to irradiation
  from the star \citep{Nayakshin10c} can result in (a) a smallish solid core and
  a complete gas envelope removal -- a terrestrial planet; (b) a massive solid
  core, with most of the gas removed -- a Uranus-like planet; (c) a partial
  envelope removal leaves a gas giant planet like Jupiter or Saturn.  For (b),
  an internal energy release due to a massive core formation removes the
  envelope \citep{HW75,Nayakshin10b}.
\end{itemize}

In contrast to the CA model, the TD scheme cannot work without a massive outer
$R\simgt $ tens to a hundred AU region of the disc. The elements (3,4) from an
earlier 1960s scenario for terrestrial planet formation
\citep{McCrea60,McCreaWilliams65} were rejected by \cite{DW75} because step (1)
is not possible in the inner Solar System. Similarly, the giant disc
instability \citep{Kuiper51,Boss98} cannot operate at $R\sim 5$ AU to make
Jupiter \citep{Rafikov05}.  It is therefore the proper placement of step (1)
into the outer reaches of the Solar System and then the introduction of the radial
migration (step 2) that makes this model physically viable. 

\cite{Nayakshin10d} suggested that, as a bonus, the new hypothesis resolves an
old mystery of the Solar System: the mainly coherent and prograde rotation of
planets, which is unexpected in the CA framework since the planets are built
by randomly oriented impacts. Note, however, that \cite{JohansenLacerda10}
shows that accretion of pebble-sized grains onto a planetary core could
provide another explanation for the observed planetary spins.

It is also not impossible \citep{Nayakshin10c} that both the TD and the CA
processes operate to sculpture the planetary systems we observe: the first in
the early, gas-rich but short ($t\simlt 10^5$ yrs) embedded period \citep{VB10},
and the second in the latter, much more quiescent phase $t\simlt$ a few
Myrs. In such a hybrid model the CA would kick-start with the benefit of the
massive terrestrial cores pre-assembled in the early TD phase.

\section{Retaining small solids}\label{sec:compositions}

\begin{figure}
\centerline{\psfig{file=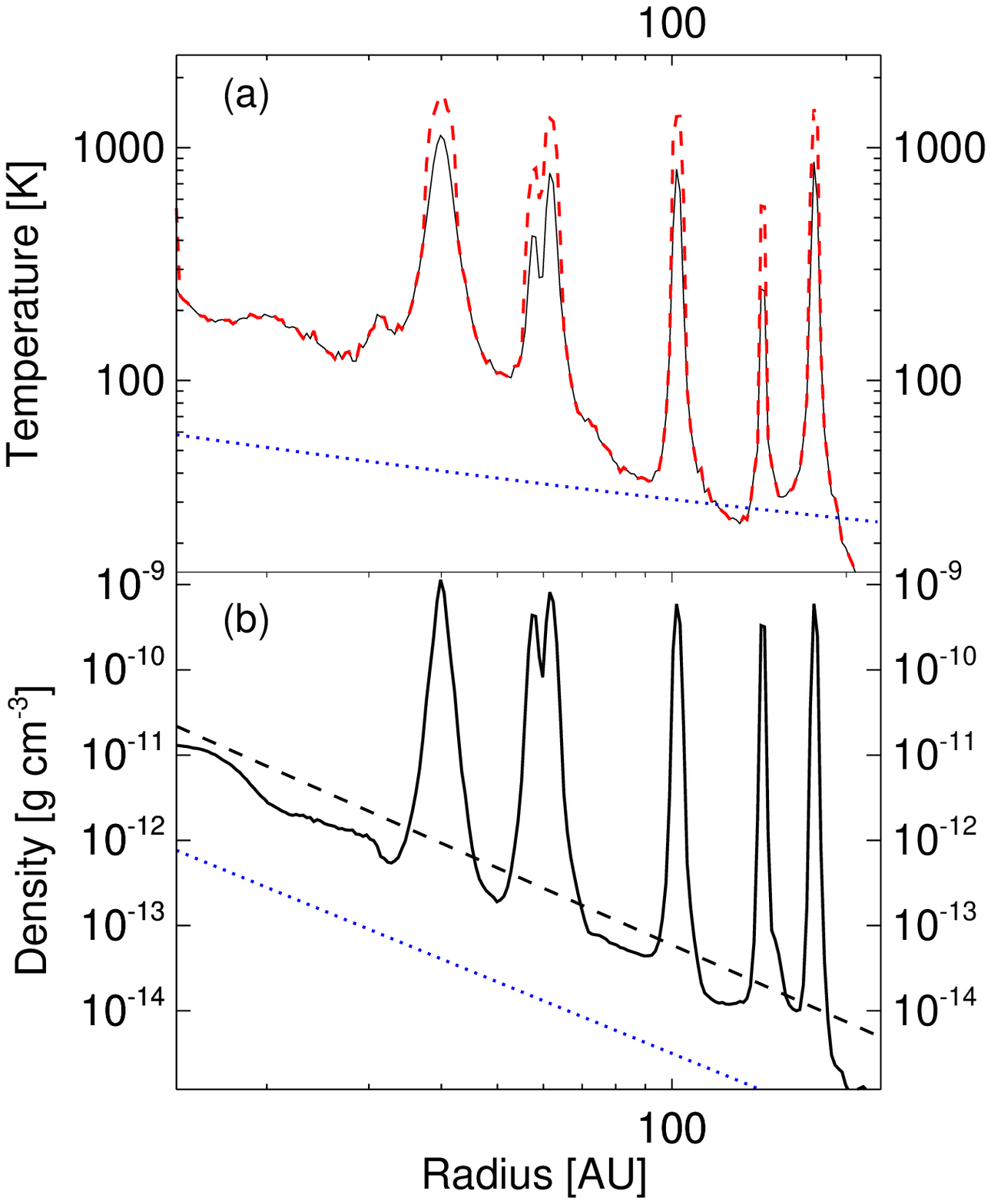,width=0.5\textwidth,angle=0}}
\label{fig:DT}
\caption{Solid curves: gas density ({\bf b}) and temperature ({\bf a})
  averaged on annuli for numerical simulation of a gas disc presented in
  \protect\cite{ChaNayakshin10}.  The red curve shows the maximum temperature found
  inside the clumps. The blue dotted curves show the corresponding temperature
  and density profiles in the standard picture of planet formation
  \citep{CG97}. The need for radial transfer of solids by factors of ten or
  more is obvious \citep[][]{Gail01,WoodenEtal07} in that theory. In contrast,
  in the TD hypothesis, the outer disc has both hot and cold regions. Mixing
  of solids produced in these two components may yield a better explanation
  for the observed composition of comets.}
\end{figure}

Although our arguments can be made completely analytical, simulations of
\cite{ChaNayakshin10} illustrate our model here. In the simulations, evolution
of a massive 0.4 $\msun$ gas disc around a 0.6 $\msun$ proto-star was followed
for about 6000 years. The massive disc becomes gravitationally unstable,
develops spiral arms, which then fragment into clumps. The black solid curve
in Figure \ref{fig:DT}a shows the annuli-averaged and density weighted gas temperature
from the simulation, defined as $<\rho T>/<\rho>$, whereas the solid curve in
Figure \ref{fig:DT}b shows the corresponding density profile, $<\rho>$, as a function of
radius $R$. The temperature and the density spikes correspond to the GEs in
the simulation. To emphasise that even higher temperatures are present in the
centres of the gas clumps, the red dashed curve in Figure \ref{fig:DT}a shows the same as
the black curve except for regions where density exceeds $\rho > 10^{-10}$ g
cm$^{-3}$, where the red curve shows the maximum temperature of the gas inside
those regions.

The black dashed curve in Figure \ref{fig:DT}b shows the tidal density of the disc,
$\rho_t = M_*/(2\pi R^3)$. Density of a disc marginally stable to the
gravitational instability would follow the dashed curve. The ``ambient'' disc,
i.e., the disc between the gas clumps, has a density lower than $\rho_t$ and
is also very cold, as expected. This confirms the two-phase division of the
outer disc suggested above.

A GE disruption should release gas and small solids with it back into the cold
disc.  Figure 2 shows the dust sedimentation time scale as a function of dust
particle radius, $a$, for GEs at three different ages from birth, $10^3$,
$10^4$ and $10^5$ years (dotted, solid and dash-triple-dotted,
respectively) \citep{Nayakshin10c}. The grains are assumed to be located at
$r_d$, set to equal exactly the half radius of the embryo, $R_{\rm GE}$, from
the embryo centre, but the results are almost independent of $r_d$. For
example, the dashed curve is same as the solid one but calculated for $r_d =
0.02 R_{\mathrm{GE}}$. As the embryos are disrupted in $\sim 10^4$~to~$10^5$
years \citep{Nayakshin10c}, small $a\ll 1$ cm grains should be abundant in
the gas envelope at the moment of disruption. Rapid radiative cooling and
mixing with the cold background naturally deposits the high-T processed
materials into the cold disc.

\section{How are the gas clumps disrupted at large radii?}

In the ``bare-bone'' version of the TD model, step (4) disrupts the envelope
by tidal forces \citep{BoleyEtal10,Nayakshin10c} or due to stellar irradiation
\citep{Nayakshin10c}. Both of these effects are weak in the outer disc, e.g.,
beyond $\sim 10$ AU. What could disrupt the GE there, and is there any
evidence for such disruption(s) in the Solar System?

As early as 35 years ago, \cite{HW75} suggested that the massive core
formation in Uranus and Neptune evaporated most of their hydrogen envelopes.
The idea here is that the energy due to core formation is trapped inside the
optically thick embryo, making it expand to sizes much larger than is expected
in the simple analytical model of the gas clumps that do not take into account
the energy release by core formation \citep{Nayakshin10c}. A more extended gas
envelope is then much easier to disrupt, even at tens of AU. 

To appreciate the argument, compare the binding energy of the solid core with
that of the GE.  We expect the core of high-Z elements to have a density
$\rho_c \sim $ a few g cm$^{-3}$. The radial size of the solid core, $R_{\mathrm
  core} \sim (3M_{\mathrm core}/4\pi\rho_c)^{1/3}$. The binding energy of the
solid core is
\begin{equation}
E_{\mathrm{bind,c}}\sim \frac{3}{5}\frac{GM_{\mathrm{core}}^{2}}{R_{\mathrm{%
core}}}\approx 10^{41}\;\hbox{erg}\;\left( \frac{M_{\mathrm{c}}}{10{\,%
\mathrm{M}_{\oplus }}}\right) ^{5/3}\;.  \label{ebind_p}
\end{equation}%
Let us now look at the gas clump itself.  The clump radius
$R_{\mathrm{GE}}\approx 0.8$ AU at the age of $t=10^{4}$ years, independently
of its mass \citep{Nayakshin10c}, $M_{\mathrm{GE}}$.  Thus, the GE binding
energy at that age is
\begin{equation}
E_{\mathrm{bind,GE}}\sim \frac{3}{10}\frac{GM_{\mathrm{GE}}^{2}}{R_{\mathrm{%
GE}}}\approx 10^{41}\;\hbox{erg}\;\left( \frac{M_{\mathrm{GE}}}{3M_{J}}%
\right) ^{2}\;.  \label{ebind_numb}
\end{equation}%
The two are comparable for $M_{\mathrm{core}}\sim 10{\,\mathrm{M}_{\oplus
}}$. Radiation hydrodynamics simulations confirm such internal disruption
events: the run labelled M$0\alpha 3$ in \cite{Nayakshin10b} made a $\sim 20{%
  \,\mathrm{M}_{\oplus }}$ solid core that unbound all but $0.03{\,\mathrm{M}%
  _{\oplus }}$ of the gaseous material of the original $10M_{J}$ gas clump.

If our model is right, then the outer $\sim$ tens of AU Solar System must
have produced at least one ``naked'' or almost so core as massive as $10 {\,%
\mathrm{M}_{\oplus}}$. There are actually two -- Uranus and Neptune with
core masses of $\sim 13$ and $15{\,\mathrm{M}_{\oplus}}$, respectively.

The self-disruption of GEs at tens or hundreds of AU is potentially
observationally testable in exoplanetary systems. The Core Accretion model %
\citep{PollackEtal96} is unlikely to produce $\lower.5ex\hbox{\gtsima}10{\,%
\mathrm{M}_{\oplus }}$ solid cores that far out; the conventional disc
instability model \citep{Boss98,BoleyEtal06} would result in massive
gas-dominated giant planets. Thus Super-Earth to Saturn mass planets on
orbits of moderate eccentricity, if found at semi-major distances of $\lower%
.5ex\hbox{\gtsima}$ tens of AU, may be sign-posts of the gas clump
self-destruction events.

Another implication of our picture is that the composition of comets and
Neptune/Uranus may be related to some degree. The crystalline materials found
in comets are materials that did not contribute to the building of the gas
giant planets. Estimates above show that forming solid cores is absolutely
essential to the release of the high-T processed materials back into the cold
disc. For the release to occur, the cores must be as massive as these outer
planets, and so may have used up a significant fraction of the solids
originally present in the GEs. Therefore materials in comets that came from
the same GE may be deficient in materials/elements abundant in Uranus and
Neptune.

\section{Implications for the origin of other high-temperature minerals in the
Solar system}\label{sec:other}

Other types of materials requiring high-temperature processes are chondrules
and Calcium Aluminium-rich Inclusions (CAIs). Chondrules are igneous-textured,
mm-size particles, composed mainly of olivine and low-Ca pyroxene set in a
feldspathic or glassy matrix \citep[e.g.,][]{ScottKrot05}.  They are a major
constituent of most chondrite groups (e.g., $\sim 80$\% of ordinary
chondrites). The origin of chondrules is controversial, but in general they
are believed to have formed as rapidly cooling molten silicate droplets.  The
maximum temperatures are taken to be approximately around the liquidus
temperatures of 1200-1500 K. The cooling rates remain uncertain for the range
of different textural types but it is thought that if chondrules had been
molten for more than a few minutes they would not have preserved the sort of
volatile abundances that they often contain \citep{YuH98}.

CAIs are the light-coloured inclusions commonly found in carbonaceous
chondrites. CAIs are more refractory-rich than chondrules. Their shapes are
less regular, while common chondrules are more uniformly
spherical. Radiometric dating using the $^{26}$Al~$-\; ^{26}$Mg chronometer suggests that
chondrules started to form $\approx 2$ Ma after CAIs \citep{McKeeganDavis05}. A
Pb-Pb absolute age for CAI formation is 4,567 Ma \citep{AmelinEtal02}.

Thus CAIs are considered to predate chondrules. Numerous chondrule and CAI
formation models include nebular shocks \citep{Cassen96}, lightning
\citep{DC00}, jets from near the proto-Sun \citep{ShuEtal01} and impacts
\citep{BridgesEtal98}. No one model is universally accepted but the impact
models have the advantage of producing abundant chondrules
\citep[e.g.,][]{ScottKrot05}. \cite{HS06} used the likely
abundance of $^{26}$Al shortly after CAI formation in the early Solar System
to show that nebular dust which rapidly accreted into $\sim 60$~km, or larger,
planetesimals would start melting. Disruption of these planetesimals by impact
would cause the sprays of melt droplets now seen preserved in chondrites.

The GEs are present only in the early ``embedded'' stage of star
formation\citep{VB10,Nayakshin10c}, which is likely to last $t\simlt 10^5$
yrs. If this is true, and if the inferred age difference between the CAIs and
chondrules is real, then GEs are likely to be dispersed or become very massive
giant planets by the time of chondrule formation. 

If we assume that formation of CAIs is co-eval with the early GE-rich epoch of
star and planet formation, then one may question whether GEs have anything to
do with CAIs.  We believe such a view is attractive because the temperatures
near the solid core inside the gas embryos may \citep{Nayakshin10a} reach
1500-2000 K, e.g., high enough formation of CAIs. Vigorous convection near the
solid core \citep{HS08,Nayakshin10b} probably drive strong shocks, which might
be one way of producing CAIs. One-dimensional simulations of
\cite{Nayakshin10b} also show melting/re-forming cycles for grains in some
cases, e.g., see the right panel of his Fig. 8, the simulation M2$\alpha
$4. Physically, the cycles result from a negative feedback loop. The accretion
luminosity of the solid core increases as grains increase in size. However,
this causes the inner GE regions to heat up, melting the grains. As grains
become smaller, the rate of their accretion onto the solid core drops, and
hence the luminosity of the core drops as well. The inner region cools down
and the grains start growing again, repeating the cycle. Thus, in the TD
model, this might explain the presence of CAIs with original sizes up to
$\simlt$ cm being found in comets. There is evidence for this from the Comet
Wild2 analyses and from IDPs (interplanetary dust particles).


\del{
Simon et al. 2008 described a CAI from Track 25 of the \textit{Stardust}
samples with a refractory mineral assemblage (perovskite CaTiO3, melilite
(CaNa)2(AlMgFe)(SiAl)2O7, spinel MgAl2O4, Al-rich pyroxene (CaMg(SiAl)O3 )
that is characteristic of CAIs from the CV3, CH, CO and CM2 carbonaceous
chondrites. Both oxygen isotopic compositions , which are 16O-rich and
fall along the Carbonaceous Chondrite Anhydrous Minerals line on a 3-isotope
plot, are indistinguishable from asteroidal CAI material. Chondrule
fragments may also have been identified in Comet Wild2. Butterworth et al.
2010 showed an olivine, pyroxene, Cr-oxide assemblage from Track 74 with
similarities to Type II chondrules. Bridges et al. 2010 described an
Al-rich diopside, pigeonite, with minor enstatite, olivine terminal grain
from Track 154 which is similar to Al-rich chondrules from carbonaceous
chondrites. Fragmentation of grains during capture by the aerogel collectors
at a closing velocity of 6 kms-1 means that accurate sizes of the original
high temperature phases in the Wild2 coma are not known. IDPs also have
some refractory mineral assemblages, having CAI-type fragments. About 55\%
of chondritic IDPs are crystalline and largely high temperature e.g.
composed of olivine and pyroxene (Ogliore et al. 2010). 
In addition to the high temperature mineralogy of cometary material, there
is also evidence for liquid water. Hydrous phyllosilicates
(clay/serpentine) have been identified in both the plume of 9P/Temple1
(Lisse et al. 2006) and also within IDPs (Rietmeijer, 2008). This must
have required temperatures to have been maintained, presumably on the
cometary nuclei, at \symbol{126}150-500oC for protracted periods of time. 
}

\begin{figure}
\centerline{\psfig{file=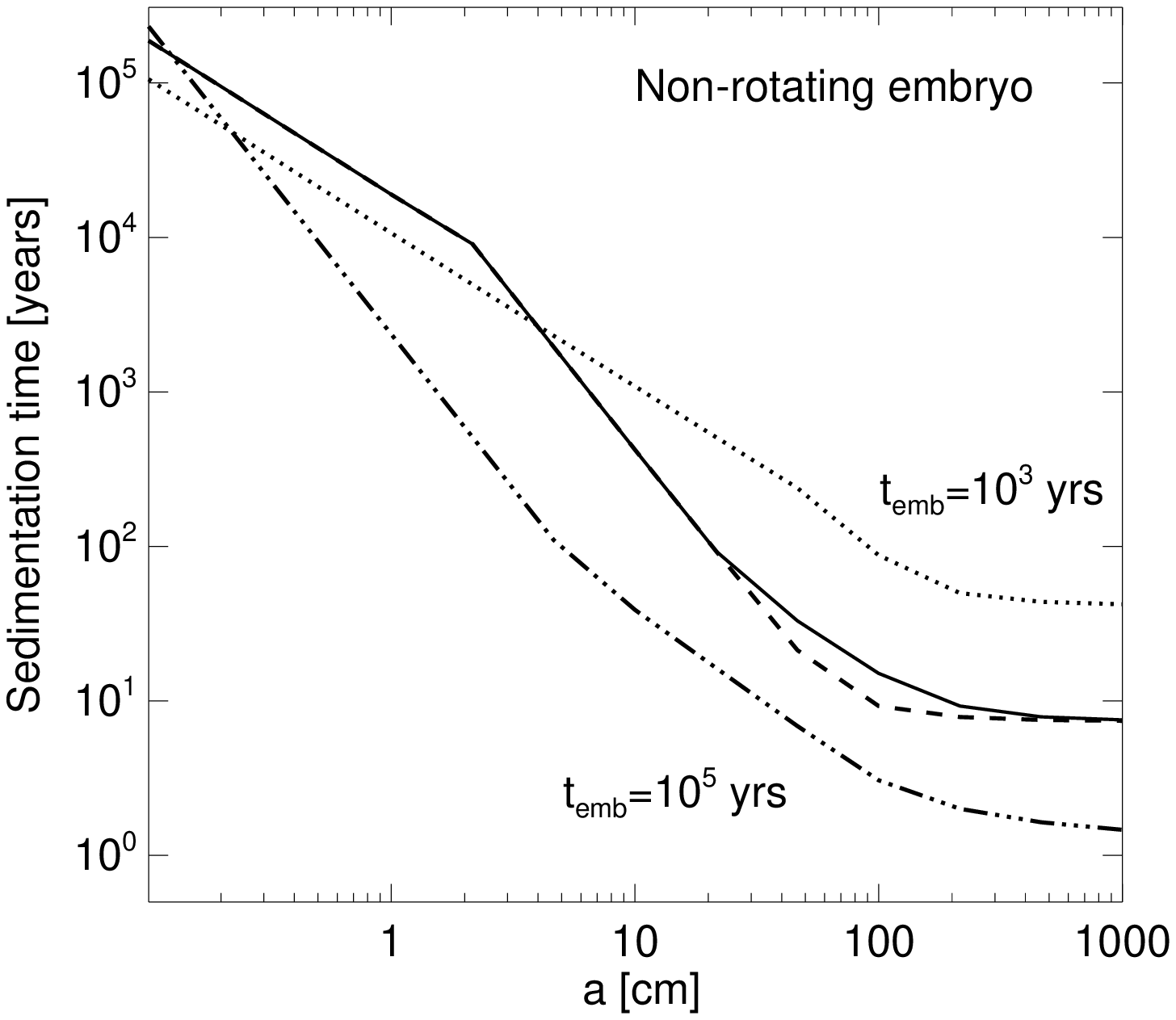,width=0.5\textwidth,angle=0}}
\caption{Sedimentation time scales for grains versus their size, $a$. All
the curves are calculated assuming the $M_{\mathrm{GE}}=10 M_J$ gas embryo
according to the model of \protect\cite{Nayakshin10c}, but at three
different embryo ages: $t=10^3$, $10^4$ and $10^5$ years, for the dotted,
the solid and the triple-dot-dashed curves, respectively. For these curves,
the grains are located at half the gas clump radius, whereas the dashed
curve shows same as the solid curve but the grains are located at 0.02 the
clump radius. The results are thus largely independent of the grain location
inside the embryo. Grains smaller than a few mm to a few cm may remain
suspended inside the embryo for a long time due to the gas drag forces. If
the embryo is disrupted, these grains are released into the surrounding cold
disc.}
\label{fig:tsed}
\end{figure}

\section{Conclusions}

We have argued for an entirely different origin of the puzzling comet
compositions. Instead of assuming that hot $T\sim 1500$ K regions needed for
thermal processing of hot minerals are located in the inner ($R\sim 1$ AU)
Solar System, we identify them with the massive and appropriately hot gas
clumps inside of which planets are born in the Tidal Downsizing hypothesis for
planet formation. In the latter, the clumps are born at radii of many tens to
hundreds of AU, and migrate inwards due to disc torques. The clumps are hot due
to their self-gravity, and due to contraction caused by radiative cooling. We
showed that small $\ll cm$ sized solids are suspended inside the gas clumps
and ought to be released back into surrounding cold disc if the clump is
disrupted. Disruption of the clumps in the outer Solar System requires a rapid
formation of massive $M > 10 \mearth$ cores inside the gas clumps, which puffs
up the gaseous envelope to the point of its removal. We tentatively identify
Uranus and Neptune as two such cores that could have disrupted their gas
embryos and contributed to building the comets in the Solar System.

There may be chemical signatures of a casual link between compositions of
comets and the cores of the icy giants confirming (or rejecting) our model,
perhaps testable with future results from the {\it Rosetta} mission. We also
note that Tidal Downsizing hypothesis predicts massive solid cores (tens of
Earth masses, e.g., planets like Uranus and Neptune and possibly more massive)
and Saturn-mass planets on semi-circular orbits at tens to hundreds of AU from
the parent stars. Such planets are unlikely to be born in either the Core
Accretion picture, where the core formation time is too long at $R\sim 100
AU$, or in the disc instability model of \citep[e.g.,][]{Boss98} where the
mass of the fragment is much more likely to exceed that of Jupiter
\citep[e.g.,][]{BoleyEtal10}. This model-differentiating prediction may be
testable with future observations of exoplanetary systems.

\section*{Acknowledgments}

The authors acknowledge illuminating discussions with and comments on the
draft by Richard Alexander. Theoretical astrophysics research and cometary
research at the University of Leicester are supported by STFC rolling grants.


\begin{thebibliography}{44}
\expandafter\ifx\csname natexlab\endcsname\relax\def\natexlab#1{#1}\fi

\bibitem[{Amelin} et~al.(2002){Amelin}, {Krot}, {Hutcheon} \&
  {Ulyanov}]{AmelinEtal02}
{Amelin} Y., {Krot} A.~N., {Hutcheon} I.~D., {Ulyanov} A.~A., 2002, Science,
  297, 1678

\bibitem[{Boley} et~al.(2010){Boley}, {Hayfield}, {Mayer} \&
  {Durisen}]{BoleyEtal10}
{Boley} A.~C., {Hayfield} T., {Mayer} L., {Durisen} R.~H., 2010, Icarus, 207,
  509

\bibitem[{Boley} et~al.(2006){Boley}, {Mej{\'{\i}}a}, {Durisen}, {Cai},
  {Pickett} \& {D'Alessio}]{BoleyEtal06}
{Boley} A.~C., {Mej{\'{\i}}a} A.~C., {Durisen} R.~H., {Cai} K., {Pickett}
  M.~K., {D'Alessio} P., 2006, \apj, 651, 517

\bibitem[{Boss}(1998)]{Boss98}
{Boss} A.~P., 1998, \apj, 503, 923

\bibitem[{Boss} et~al.(2002){Boss}, {Wetherill} \& {Haghighipour}]{BossEtal02}
{Boss} A.~P., {Wetherill} G.~W., {Haghighipour} N., 2002, Icarus, 156, 291

\bibitem[{Bridges} et~al.(1998){Bridges}, {Franchi}, {Hutchison}, {Sexton} \&
  {Pillinger}]{BridgesEtal98}
{Bridges} J.~C., {Franchi} I.~A., {Hutchison} R., {Sexton} A.~S., {Pillinger}
  C.~T., 1998, Earth and Planetary Science Letters, 155, 183

\bibitem[{Cassen}(1996)]{Cassen96}
{Cassen} P., 1996, in { Chondrules and the Protoplanetary Disk\/}, edited by
  {R.~Hewins, R.~Jones, \& E.~Scott},  21--28

\bibitem[{Cha} \& {Nayakshin}(2010)]{ChaNayakshin10}
{Cha} S.-H., {Nayakshin} S., 2010, {submitted to MNRAS Letters}

\bibitem[{Chiang} \& {Goldreich}(1997)]{CG97}
{Chiang} E.~I., {Goldreich} P., 1997, \apj, 490, 368

\bibitem[{Desch} \& {Cuzzi}(2000)]{DC00}
{Desch} S.~J., {Cuzzi} J.~N., 2000, \icarus, 143, 87

\bibitem[{Donnison} \& {Williams}(1975)]{DW75}
{Donnison} J.~R., {Williams} I.~P., 1975, \mnras, 172, 257

\bibitem[{Gail}(2001)]{Gail01}
{Gail} H., 2001, \aap, 378, 192

\bibitem[{Goldreich} \& {Tremaine}(1980)]{GoldreichTremaine80}
{Goldreich} P., {Tremaine} S., 1980, \apj, 241, 425

\bibitem[{Handbury} \& {Williams}(1975)]{HW75}
{Handbury} M.~J., {Williams} I.~P., 1975, AP\&SS, 38, 29

\bibitem[{Helled} \& {Schubert}(2008)]{HS08}
{Helled} R., {Schubert} G., 2008, Icarus, 198, 156

\bibitem[{Hevey} \& {Sanders}(2006)]{HS06}
{Hevey} P.~J., {Sanders} I.~S., 2006, Meteoritics and Planetary Science, 41, 95

\bibitem[{Hughes} \& {Armitage}(2010)]{HA10}
{Hughes} A.~L.~H., {Armitage} P.~J., 2010, \apj, 719, 1633

\bibitem[{Ida} \& {Lin}(2008)]{IdaLin08}
{Ida} S., {Lin} D.~N.~C., 2008, \apj, 685, 584

\bibitem[{Johansen} \& {Lacerda}(2010)]{JohansenLacerda10}
{Johansen} A., {Lacerda} P., 2010, \mnras, 404, 475

\bibitem[{Kawakita} et~al.(2004){Kawakita}, {Watanabe}, {Furusho}
  et~al.]{KawakitaEtal04}
{Kawakita} H., {Watanabe} J., {Furusho} R., et~al., 2004, \apj, 601, 1152

\bibitem[{Kuiper}(1951)]{Kuiper51}
{Kuiper} G.~P., 1951, in { 50th Anniversary of the Yerkes Observatory and Half
  a Century of Progress in Astrophysics\/}, edited by {J.~A.~Hynek},  357--+

\bibitem[{Larson}(1969)]{Larson69}
{Larson} R.~B., 1969, \mnras, 145, 271

\bibitem[{Lin} et~al.(1996){Lin}, {Bodenheimer} \& {Richardson}]{Lin96}
{Lin} D.~N.~C., {Bodenheimer} P., {Richardson} D.~C., 1996, \nat, 380, 606

\bibitem[{McCrea}(1960)]{McCrea60}
{McCrea} W.~H., 1960, Royal Society of London Proceedings Series A, 256, 245

\bibitem[{McCrea} \& {Williams}(1965)]{McCreaWilliams65}
{McCrea} W.~H., {Williams} I.~P., 1965, Royal Society of London Proceedings
  Series A, 287, 143

\bibitem[{McKeegan} \& {Davis}(2005)]{McKeeganDavis05}
{McKeegan} K.~D., {Davis} A.~M., 2005, {Early Solar System Chronology},
  431--+, Elsevier B

\bibitem[{Nayakshin}(2010{\natexlab{a}})]{Nayakshin10e}
{Nayakshin} S., 2010{\natexlab{a}}, ArXiv e-prints

\bibitem[{Nayakshin}(2010{\natexlab{b}})]{Nayakshin10c}
{Nayakshin} S., 2010{\natexlab{b}}, \mnras, 408, L36

\bibitem[{Nayakshin}(2010{\natexlab{c}})]{Nayakshin10b}
{Nayakshin} S., 2010{\natexlab{c}}, ArXiv e-prints, 1007.4165

\bibitem[{Nayakshin}(2010{\natexlab{d}})]{Nayakshin10a}
{Nayakshin} S., 2010{\natexlab{d}}, \mnras, 408, 2381

\bibitem[{Nayakshin}(2011)]{Nayakshin10d}
{Nayakshin} S., 2011, \mnras, 410, L1

\bibitem[{Pollack} et~al.(1996){Pollack}, {Hubickyj}, {Bodenheimer},
  {Lissauer}, {Podolak} \& {Greenzweig}]{PollackEtal96}
{Pollack} J.~B., {Hubickyj} O., {Bodenheimer} P., {Lissauer} J.~J., {Podolak}
  M., {Greenzweig} Y., 1996, Icarus, 124, 62

\bibitem[{Rafikov}(2005)]{Rafikov05}
{Rafikov} R.~R., 2005, \apjl, 621, L69

\bibitem[{Rafikov}(2010)]{Rafikov10}
{Rafikov} R.~R., 2010, ArXiv e-prints

\bibitem[{Safronov}(1969)]{Safronov69}
{Safronov} V.~S., 1969, {Evoliutsiia doplanetnogo oblaka.}

\bibitem[{Safronov}(1972)]{Safronov72}
{Safronov} V.~S., 1972, {Evolution of the protoplanetary cloud and formation of
  the earth and planets.}, Jerusalem (Israel): Israel Program for Scientific
  Translations, Keter Publishing House, 212 p.

\bibitem[{Scott} \& {Krot}(2005)]{ScottKrot05}
{Scott} E.~R.~D., {Krot} A.~N., 2005, {Chondrites and their Components},
  143--+, Elsevier B

\bibitem[{Shu} et~al.(2001){Shu}, {Shang}, {Gounelle}, {Glassgold} \&
  {Lee}]{ShuEtal01}
{Shu} F.~H., {Shang} H., {Gounelle} M., {Glassgold} A.~E., {Lee} T., 2001,
  \apj, 548, 1029

\bibitem[{Vorobyov} \& {Basu}(2010)]{VB10}
{Vorobyov} E.~I., {Basu} S., 2010, ArXiv e-prints

\bibitem[{Westphal} et~al.(2009){Westphal}, {Fakra}, {Gainsforth}, {Marcus},
  {Ogliore} \& {Butterworth}]{WestphalEtal09}
{Westphal} A.~J., {Fakra} S.~C., {Gainsforth} Z., {Marcus} M.~A., {Ogliore}
  R.~C., {Butterworth} A.~L., 2009, \apj, 694, 18

\bibitem[{Wooden} et~al.(2007){Wooden}, {Desch}, {Harker}, {Gail} \&
  {Keller}]{WoodenEtal07}
{Wooden} D., {Desch} S., {Harker} D., {Gail} H., {Keller} L., 2007, Protostars
  and Planets V,  815--833

\bibitem[{Wooden} et~al.(2005){Wooden}, {Harker} \& {Brearley}]{WoodenEtal05}
{Wooden} D.~H., {Harker} D.~E., {Brearley} A.~J., 2005, in { Chondrites and the
  Protoplanetary Disk\/}, edited by {A.~N.~Krot, E.~R.~D.~Scott, \&
  B.~Reipurth}, vol. 341 of { Astronomical Society of the Pacific Conference
  Series\/},  774--+

\bibitem[{Yu} \& {Hewins}(1998)]{YuH98}
{Yu} Y., {Hewins} R.~H., 1998, Geochim. Cosmochim. Acta, 62, 159

\bibitem[{Zolensky} et~al.(2006){Zolensky}, {Zega}, {Yano}
  et~al.]{ZolenskyEtal06}
{Zolensky} M.~E., {Zega} T.~J., {Yano} H., et~al., 2006, Science, 314, 1735

\end{thebibliography}

\label{lastpage}

\end{document}